\documentclass[aps,showkeys,twocolumn,amsmath,amssymb,superscriptaddress]{revtex4}
\usepackage{graphicx}
\usepackage{dcolumn}
\usepackage{amsmath}
\usepackage{bm}
\usepackage{natbib}

\begin{document}

\title{Measuring multiple spike train synchrony}

\author{Thomas Kreuz}
\email{thomas.kreuz@fi.isc.cnr.it} \affiliation{Institute for Nonlinear Sciences, University of California, San Diego, USA} \affiliation{Istituto dei Sistemi Complessi - CNR,
Via Madonna del Piano 10, I-50019 Sesto Fiorentino, Italy}
\author{Daniel Chicharro}
\affiliation{Department of Information and Communication Technologies, Universitat Pompeu Fabra, Barcelona, Spain}
\author{Ralph G. Andrzejak}
\affiliation{Department of Information and Communication Technologies, Universitat Pompeu Fabra, Barcelona, Spain}
\author{Julie S. Haas}
\affiliation{Center for Brain Science, Harvard University, MA, USA}
\author{Henry D. I. Abarbanel}
\affiliation{Institute for Nonlinear Sciences, University of California, San Diego, USA} \affiliation{Department of Physics and Marine Physical Labaratory (Scripps Institution
of Oceanography), University of California, San Diego, USA}
\date{\today}

\begin{abstract}
    Measures of multiple spike train synchrony are essential in order to study issues such as spike timing reliability, network synchronization, and neuronal coding. These measures
    can broadly be divided in multivariate measures and averages over bivariate measures. One of the most recent bivariate approaches, the ISI-distance, employs the ratio
    of instantaneous interspike intervals (ISIs). In this study we propose two extensions of the ISI-distance, the straightforward averaged bivariate ISI-distance and the
    multivariate ISI-diversity based on the coefficient of variation. Like the original measure these extensions combine many properties desirable in applications to real data.
    In particular, they are parameter free, time scale independent, and easy to visualize in a time-resolved manner, as we illustrate with \emph{in vitro} recordings from a
    cortical neuron. Using a simulated network of Hindemarsh-Rose neurons as a controlled configuration we compare the performance of our methods in distinguishing different levels
    of multi-neuron spike train synchrony to the performance of six other previously published measures. We show and explain why the averaged bivariate measures perform better than
    the multivariate ones and why the multivariate ISI-diversity is the best performer among the multivariate methods. Finally, in a comparison against standard
    methods that rely on moving window estimates, we use single-unit monkey data to demonstrate the advantages of the instantaneous nature of our methods.
\end{abstract}


\newcommand{\abb}{\small\sf}

\maketitle
%
%
\section{\label{s:Intro} Introduction}

Estimating the degree of synchrony within a set of spike trains is a common task within two different scenarios. In the first scenario spike trains are recorded successively
from only one neuron. The most common application is to quantify the reliability of the neuronal response upon repeated presentations of the same stimulus \citep{Mainen95,
Tiesinga08}. In the second scenario spike trains are derived from simultaneous recordings of a population of neurons \citep{Gerstein01, Brown04}. A typical application for such
data is the analysis of synchronization and desynchronization in large neural networks \citep{Buzsaki04}.

Measures designed to estimate single-neuron reliability and multi-neuron synchrony are either multivariate or bivariate. Multivariate approaches evaluate all spike trains at the
same time, e.g., by measuring the normalized variance of pooled, exponentially convolved spike trains \citep{Hunter98}, or by exploiting the deviation of pooled spike train
statistics from the one obtained for a Poisson process \citep{Tiesinga04}. Bivariate measures aim at the quantification of synchrony between just two spike trains, but can also
be used in a multivariate context by defining the total synchrony as the average over all pairwise synchrony-values \citep{Kreiman00}. Prominent examples are the cost-based
distance introduced in \citet{Victor96}, the Euclidean distance proposed in \citet{VanRossum01}, cross correlation of spike trains after filtering \citep{Haas02, S_Schreiber03},
and event synchronization \citep{QuianQuiroga02b}.

Most recently, the bivariate ISI-distance has been introduced as a simple approach that extracts information from the interspike intervals (ISIs) by evaluating the ratio of
instantaneous firing rates \citep{Kreuz07c}. The ISI-distance combines a variety of properties that make it ideally suited for applications to real data. In particular, it
serves as an excellent means to visualize the occurrence of spiking patterns. In contrast to most of the other measures that rely on a parameter determining the time scale, the
ISI-distance does not depend on any time scales and is thus free of parameters. In a comparison with previously published approaches on spike trains extracted from a simulated
Hindemarsh-Rose network, the ISI-distance matched the performance of the best time scale optimized measure \citep{Kreuz07c}.

In this study we propose two extensions of the ISI-distance, the straightforward averaged bivariate ISI-distance and the multivariate ISI-diversity based on the coefficient of
variation. Both of these extensions inherit the fundamental properties of the measure they are based on. In particular, they are parameter free, time scale independent and easy
to visualize in a time-resolved manner. We apply the two measures to \emph{in vitro} recordings of a cortical cell from a rat and show that they serve as an excellent means to
track spiking patterns in multiple spike trains.

In order to better characterize the two extensions of the ISI-distance, we consider them in the context of the groups of averaged bivariate and multivariate measures to which
they each belong. In particular, we compare their performance in distinguishing different levels of multi-neuron spike train synchrony to the performance of six other previously
published averaged bivariate and multivariate measures. As a controlled configuration we use a simulated network of Hindemarsh-Rose neurons with predefined clustering
\citep{Morelli05, Kreuz07c}. In this comparison, the averaged bivariate measures perform better than the multivariate measures. Finally, we present a typical application using
single-unit data from a visual attention task in a macaque monkey \citep{Mitchell07}. Comparing the ISI-measures to standard methods that rely on moving window estimates, we
stress the advantages of their instantaneous nature of our proposed methods.

The remainder of the paper is organized as follows: In Section \ref{s:ISI-distances} we present a more detailed description of the original bivariate ISI-distance (Section
\ref{ss:Bivariate-ISI-distance}), the averaged bivariate ISI-distance (Section \ref{ss:Averaged-ISI-distance}), and the multivariate ISI-diversity (Section
\ref{ss:Multivariate-ISI-distance}). Furthermore, we explain how to evaluate the significance of the values obtained in order to study properties of neuronal coding (Section
\ref{ss:Neuronal coding}). All of these measures are applied to \emph{in vitro} recordings from a cortical neuron in Section \ref{s:Illustration-Real-Data}. Subsequently, in
Section \ref{s:Comparison-HR} a performance comparison is carried out on a network of Hindemarsh-Rose neurons. Finally, in Section \ref{s:Example-Jude} a typical field data
application is shown before the conclusions are drawn in Section \ref{s:Discussion}. In the Appendix we derive analytical results for Poisson processes in Section
\ref{App-s:Theory}, introduce the \emph{in vitro} recordings of cortical cells, the simulated Hindemarsh-Rose time series, and the \emph{in vivo} monkey data in Sections
\ref{App-ss:Data-Stellate-Cells}, \ref{App-ss:Data-Hindemarsh-Rose}, and \ref{App-ss:Jude-Data} respectively, and describe the previously published approaches to measure spike
train synchrony in Section \ref{App-s:Existing-Methods}.
%
%

\section{\label{s:ISI-distances} The bivariate ISI-distance and its extensions}

We denote with $\{t_i^n\} = {t_1^n,...,t_{M_n}^n}$ the spike times and with $M_n$ the number of spikes for neuron $n$ with $n = 1,...,N$. As a first step common to all the
following instantaneous measures for each neuron the value of the current interspike interval is assigned to each time instant\footnote{There are several ways of dealing with
the recording's temporal edges. Here we take into account only the interval that is covered by all individual spike trains. For other options and more general information on the
implementation as well as the Matlab source code for calculating and visualizing the ISI-distance and its extensions please refer to
http://inls.ucsd.edu/$\sim$kreuz/Source-Code/Spike-Sync.html.}
\begin{equation} \label{eq:ISI}
    x_{\mathrm {ISI}}^n (t) = \min(t_i^n | t_i^n > t) - \max(t_i^n | t_i^n < t) \quad t_1^n < t < t_{M_n}^n.
\end{equation}
From the instantaneous ISIs the average instantaneous rate can be defined as
\begin{equation} \label{eq:ISI-Rate}
    R (t) = \frac{1}{N} \sum_{n=1}^N \frac{1}{x_{\mathrm {ISI}}^n (t)}.
\end{equation}

\subsection{\label{ss:Bivariate-ISI-distance} Bivariate ISI-distance}

To define a time-resolved, symmetric, and scale-invariant measure of the relative firing rate pattern \citep{Kreuz07c} we take the instantaneous ISI-ratio between $x_{\mathrm
{ISI}}^1$ and $x_{\mathrm {ISI}}^2$, and normalize according to:
\begin{equation} \label{eq:ISI-Ratio}
    I_{12} (t) = \begin{cases}
           x_{\mathrm {ISI}}^1 (t) / x_{\mathrm {ISI}}^2 (t) - 1 & {\rm if} ~~ x_{\mathrm {ISI}}^1 (t) \leq x_{\mathrm {ISI}}^2 (t) \cr
                      - (x_{\mathrm {ISI}}^2 (t) / x_{\mathrm {ISI}}^1 (t) -1)     & {\rm otherwise}.
                  \end{cases}
\end{equation}
This quantity becomes $0$ for identical ISI in the two spike trains, and approaches $-1$ and $1$, respectively, if the first (second) spike train is much faster than the other.

Finally, the absolute ISI-distance is integrated over time:
\begin{equation} \label{eq:ISI-distance}
    D_I = \frac{1}{T} \int_{t=0}^T dt | I_{12} (t) |.
\end{equation}
Identical spike trains yield a value of zero if there is no phase lag between the spike trains (for ways to deal with non-zero phase lags please confer the discussion in Section
\ref{s:Discussion}). Irrespective of any phase lag the value zero is also obtained for periodic spike trains with the same period (i.e., all ISIs of the two spike trains have
the same length). For the more general case of periodic spike trains with periods $p$ and $q$ (with $p \leq q$) a value of
\begin{equation} \label{eq:ISI-periodic}
    D_I = 1 - p/q
\end{equation}
is obtained. For the special case that $p$ and $q$ are integer values, this is the classical $p:q$ synchronization.

\subsection{\label{ss:Averaged-ISI-distance} Averaged bivariate ISI-distance}

We again start by assigning the ISI for each spike train $\{t_i^n\}$  with $n = 1,...,N$ as in Eq. \ref{eq:ISI} and proceed by calculating the instantaneous average $A (t)$ over
all pairwise absolute ISI-ratios $|I_{mn} (t)|$ (cf. Eq. \ref{eq:ISI-Ratio})
\begin{equation} \label{eq:Instantaneous-Average}
    A (t) = \frac{1}{N(N-1)/2}\sum_{n=1}^N \sum_{m=n+1}^N | I_{mn}(t) |
\end{equation}
Averaging over time yields
\begin{equation} \label{eq:Multi-ISI-distance}
    D_I^a = \frac{1}{T} \int_{t=0}^T dt A (t).
\end{equation}

The same kind of time-resolved visualization as in the bivariate case is possible, because the two averages commute. Instead of averaging over pairwise ISI-distances, which
themselves are obtained by averaging over time, we can first average over pairwise instantaneous ISI-ratios and then average over time.

\subsection{\label{ss:Multivariate-ISI-distance} Multivariate ISI-diversity based on coefficient of variation}

We derive a multivariate measure by taking the instantaneous coefficient of variation taken across all neurons $n = 1,...,N$ at any given instant in time
\begin{equation} \label{eq:ISI-CV}
    C_V (t) = \frac{\sigma(x_{\mathrm {ISI}}^n (t))}{\langle x_{\mathrm {ISI}}^n (t) \rangle_n}
\end{equation}
and again integrate over time:
\begin{equation} \label{eq:Multi-Intra-ISI-distance}
    D_I^m = \frac{1}{T} \int_{t=0}^T dt C_V(t).
\end{equation}
The coefficient of variation is probably the most widely used measure of variability. Since it relates the standard deviation to the mean (and therefore is dimensionless), it
allows for broader comparisons also between data sets with different units or different means. For identical spike trains and for periodic spike trains with the same period it
obtains the same lower bound of zero as $D_I$ and $D_I^a$ but unlike them it lacks an upper bound.

The coefficient of variation can be very sensitive to outliers in the instantaneous ISI distribution. Since this distribution is bounded below by zero and typically skewed
towards larger values, we first log-transform the data and then calculate the back-transformed coefficient of variance according to \citet{Hopkins00}
\begin{equation} \label{eq:ISI-CV2}
    C_V^{log} (t) = \langle \ln(x_{\mathrm {ISI}}^n (t)) \rangle_n \exp(\frac{\sigma(\ln(x_{\mathrm {ISI}}^n (t)))}{\langle \ln(x_{\mathrm {ISI}}^n (t)) \rangle_n})-1.
\end{equation}

\subsection{\label{ss:Neuronal coding} ISI-distances and neuronal coding}

The ISI-distance and its extensions provided here, as well as the other spike train distances, are not designed to directly address the question of what is coded by the neurons
and in which way this is done, as for example is the case for many approaches that rely on spike triggering \citep{Schwartz06}. However, they allow examining properties such as
the reliability of a neuron to the same stimulus or similarity across a population in a single presentation of a stimulus which can help to infer and characterize the neuronal
response. For that purpose it is important to compare the values obtained to a baseline level indicating the results that would be obtained if different spike trains are assumed
to be independent. One of the most common reference models in spike train analysis is the Poisson process, an example for a renewal process for which at each time instant the
probability of a spike is independent of the occurrence of previous spikes. In Appendix \ref{App-s:Theory} we investigate the case of multiple Poisson processes and derive
analytical results for the ISI-distance $D_I$, the averaged bivariate ISI-distance $D_I^a$ and the ISI-diversity $D_I^m$ all of which can be used as benchmarks for this type of
random spike trains.

\section{\label{s:Illustration-Real-Data} Illustration of the ISI-measures using \emph{in vitro} recordings from a cortical cell}

\begin{figure}
    \includegraphics[width=85mm]{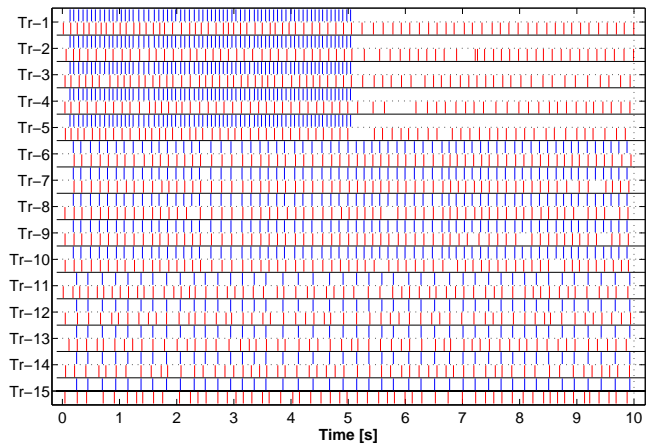}
    \caption{\abb\label{fig:RealData} (color online) Input and output spike trains recorded from a cortical cell. The same input patterns were delivered with three
        different firing rates that were each repeated five times (Trials 1-5: faster input, Trials 6-10: matched input, Trials 11-15: slower input).
        The input is shown in blue and the output in red.}
\end{figure}

\begin{figure}
    \includegraphics[width=85mm]{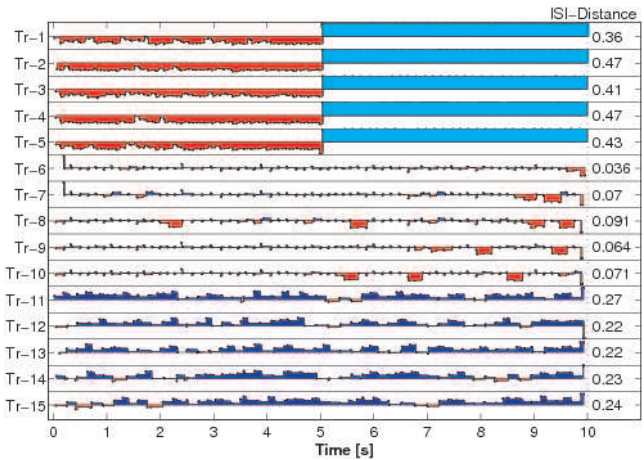}
    \caption{\abb\label{fig:RealData-Pairwise-ISI-distances} (color online) Pairwise input-output ISI-ratios for the fifteen trials shown in Fig.~\ref{fig:RealData}.
        The ISI-distances $D_I$, defined as the average over the absolute values of the ISI-ratios (Eq. \ref{eq:ISI-distance}), are shown on the right side of each trial.
        For the fast inputs (Trials 1-5) this is the ISI-distance averaged over the first half of the recording only, the part after the last input spike is ignored.}
\end{figure}

In an example application, the averaged bivariate ISI-distance $D_I^a$ and the multivariate ISI-diversity $D_I^m$ are illustrated using \emph{in vitro} whole-cell recordings
taken from a cortical cell of a young Long-Evans rat. For a description of the data see Appendix \ref{App-ss:Data-Stellate-Cells}. Sets of synaptic inputs were delivered to a
spiking neuron using three different input firing rates with five repetitions each. All recordings including the synaptic input trains are shown in Fig.~\ref{fig:RealData} and
their pairwise ISI-ratios (Eq. \ref{eq:ISI-Ratio}) are depicted in Fig.~\ref{fig:RealData-Pairwise-ISI-distances}.

In the first five trials the input is too fast for the neuron to follow, while in the last five trials the neuron does not wait for the slower input and maintains its own,
faster firing rate. In trials $6-10$ the input firing rate matches the neuron's own rate, such that except for small deviations $1:1$ synchronization can be observed. The mean
ISI-distance for the first five trials (first half only) is $D_I = 0.43$ while it is $D_I = 0.24$ for the last five trials. However, the lowest distances are obtained for the
matched inputs, for which all distances are smaller than $0.1$ with a mean distance of $D_I = 0.066$. This resonance-like behavior is consistent with results recently reported
in \citet{Haas09}.

\begin{figure}
    \includegraphics[width=85mm]{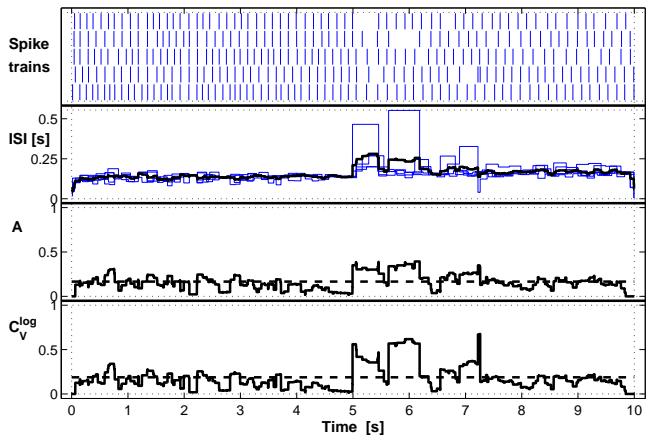}
    \caption{\abb\label{fig:PA-CV-ISI-distance-RealData-1} (color online) Relative firing patterns for output spike trains recorded using faster input (red traces $1-5$ in
        Fig.~\ref{fig:RealData}). On top we depict the spike trains, followed by the instantaneous ISI-values according to Eq. \ref{eq:ISI} and their instantaneous average
        (black line). At the bottom we show the pairwise average (Eq. \ref{eq:Instantaneous-Average}) and the coefficient of variation (Eq. \ref{eq:ISI-CV}). Dotted vertical
        lines mark beginning and end of the recording; dashed lines in the lower part indicate the average values over time, respectively. These are the ISI-measures which
        for this set of spike trains are equal to $D_I^a = 0.17$ and $D_I^m = 0.19$, respectively. For the first half only both measures equal $0.13$.}
\end{figure}

\begin{figure}
    \includegraphics[width=85mm]{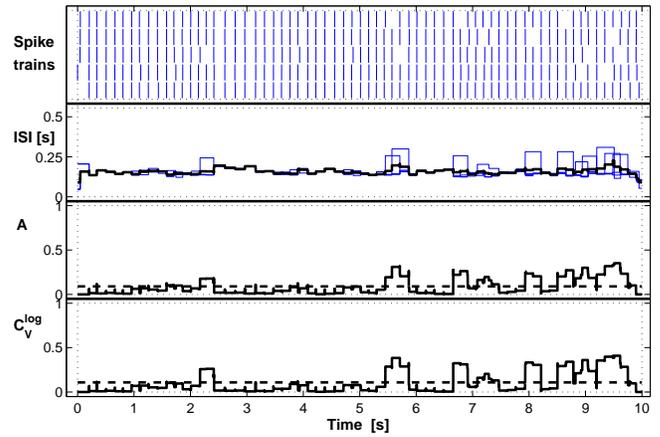}
    \caption{\abb\label{fig:PA-CV-ISI-distance-RealData-2} (color online) Same as Fig. \ref{fig:PA-CV-ISI-distance-RealData-1} but this time for the five output spike trains
        recorded during stimulation with speed-matched input (red traces $6-10$ in Fig.~\ref{fig:RealData}). For this example the ISI-measures attain the low values
        $D_I^a = 0.09$ and $D_I^m = 0.11$.}
\end{figure}

\begin{figure}
    \includegraphics[width=85mm]{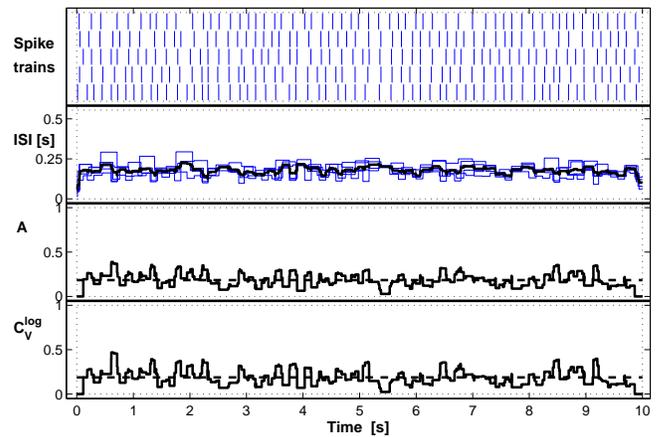}
    \caption{\abb\label{fig:PA-CV-ISI-distance-RealData-3} (color online) As Fig. \ref{fig:PA-CV-ISI-distance-RealData-1} but this time for slower inputs. Here the
        ISI-measures are $D_I^a = 0.18$ and $D_I^m = 0.19$.}
\end{figure}

Here we use the bivariate ISI-distance $D_I$ to address the input-output relation for each trial; next we employ the averaged bivariate ISI-distance $D_I^a$ and the multivariate
ISI-diversity $D_I^m$ (after log-transform, Eq. \ref{eq:ISI-CV2}) to estimate the reliability of the output across repetitions of the same input, and the dependence on the
input's temporal scaling. In Figs.~\ref{fig:PA-CV-ISI-distance-RealData-1} to ~\ref{fig:PA-CV-ISI-distance-RealData-3} we investigate spiking patterns in the output spike trains
of faster, temporally matched and slower inputs, respectively.

For the faster inputs (trials 1-5), which lasted only half the recording, the change in firing behavior at the transition can easily be tracked
(Fig.~\ref{fig:PA-CV-ISI-distance-RealData-1}). In the beginning all output spike trains are fast and regular (cf. instantaneous ISI in
Fig.~\ref{fig:PA-CV-ISI-distance-RealData-1}) and exhibit similar spiking patterns (cf. instantaneous variabilities $A$ and $C_V^{log}$ in
Fig.~\ref{fig:PA-CV-ISI-distance-RealData-1}). Immediately after the last input spike, at approximately $5$ seconds, some of the output spike trains become very irregular. This
results in high values of the instantaneous pairwise averages and the instantaneous coefficients of variation. After this transition period the variability of the spike trains
returns to the initial low values, although now at a somewhat lower rate (higher instantaneous ISIs).

Much lower ISI-measures are obtained for the matched input firing rates (trials 6-10, cf. Fig.~\ref{fig:PA-CV-ISI-distance-RealData-2}). Firing is more uniform and reliability
is higher as reflected by the low values of the instantaneous bivariate averages and the instantaneous coefficients of variation. Occasionally these time-resolved measures reach
$0$, marking periods of perfect global ISI synchronization, whereas in the second half they exhibit an increasing trend, which correctly reflects missing spikes in some of the
spike trains.

For slower input (trials 11-15) the ISI-measures yield again higher values similar to the ones for slower inputs (cf. Fig.~\ref{fig:PA-CV-ISI-distance-RealData-3}). However, for
slower inputs the high variability is spread uniformly over the whole time interval, while the high distances for faster inputs are mostly due to the large irregularities at the
termination of the input.

In summary, this qualitative analysis with the ISI-measures can confirm previous observations on the dependence of reliability on input firing rate \citep{Haas09}. Furthermore,
the time-resolved visualization used here allows for a more detailed analysis of the spiking patterns than measures that compress all the information into one final value can
achieve.

%
%
\section{\label{s:Comparison-HR} Comparison of measures using simulated Hindemarsh-Rose time series}

In this section we evaluate how well the ISI-measures and six previously published methods perform in distinguishing different levels of multi-neuron spike train synchrony. In
order to change the level of synchrony in a controlled manner, we constructed sets of spike trains from two clusters and gradually varied the relative contributions of these
clusters. This configuration is reminiscent of multi-unit recordings in which a given set of spike trains might represent different types of neurons or neurons responding to
different tasks. In the first part of this section we show that the synchrony within such a non-homogeneous neuronal population is dependent on both the total number and the
relative mixture of neurons. In the second part we define a criterion for comparing the performance of the various measures in distinguishing different sets of spike trains.

\begin{figure}
    \includegraphics[width=85mm]{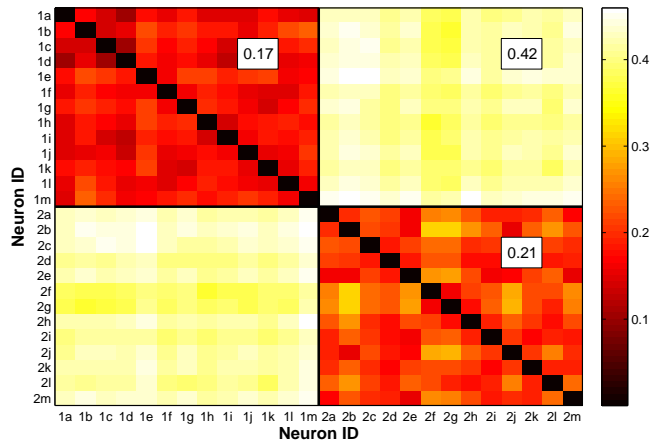}
    \caption{\abb\label{fig:Bi-ISI-Dist-Mat} (color online) Pairwise distance matrix for the bivariate ISI-distance $D_I$ applied to $26$ time series from the Hindemarsh-Rose model.
        Neurons and their spike trains are labelled by '1' and '2' depending on their affiliation to the two clusters. Black lines separate the two intra-cluster and the inter-cluster
        submatrices. Their respective mean ISI-distances are shown. Note the higher mean ISI-distance $D_I^a$ of cluster $2$ as compared to the one of cluster $1$.}
\end{figure}

The dataset consists of $26$ spike trains taken from a network of Hindmarsh-Rose neurons. It is a subsample of the data previously used in \citet{Kreuz07c}. In a simulation the
network architecture was designed such that the $26$ spike trains arise from two principal clusters with $13$ neurons each. This organization is reflected by the pairwise
distance matrix shown in Fig. \ref{fig:Bi-ISI-Dist-Mat} for the bivariate ISI-distance $D_I$. Note that the first cluster is more strongly connected than the second; the
corresponding mean ISI-distances are $\langle D_I \rangle_1 = 0.17$ and $\langle D_I \rangle_2 = 0.21$, respectively.

As a benchmark test, we compared the averaged bivariate ISI-distance $D_I^a$ (Eq. \ref{eq:Multi-ISI-distance}) and the multivariate ISI-diversity $D_I^m$ (after log-transform,
Eq. \ref{eq:ISI-CV2}) with four averaged bivariate methods and two multivariate measures. The averaged bivariate measures were comprised of the spike train distances $D_V^a$ and
$D_R^a$ introduced by \citet{Victor96} and \citet{VanRossum01}, the similarity measure $D_S^a$ proposed by \citet{S_Schreiber03} and event synchronization $D_Q^a$
\citep{QuianQuiroga02b}. Multivariate measures included the reliabilities $D_H^m$ and $D_T^m$ by \citet{Hunter98} and \citet{Tiesinga04}, respectively. Four of these measures
($D_V^a$, $D_R^a$, $D_S^a$ and $D_H^m$) rely on the choice of a parameter that sets the time scale. Following \citet{Kreuz07c} these time scales were optimized with respect to
the present task (see below). For a description of the underlying model confer to Appendix \ref{App-ss:Data-Hindemarsh-Rose}. Details about the previously published measures and
their parameters can be found in Appendix \ref{App-s:Existing-Methods}.

To evaluate how well the measures are able to distinguish different sets of spike trains, we performed $12$ different runs with the total number of spike trains increasing from
$N = 2$ to $N = 13$. In each run we varied the numbers $N_1$ and $N_2$ of spike trains from the two clusters but kept the total number $N = N_1 + N_2$ of spike trains within the
set constant. For any given $N$ we started with spike trains from cluster $2$ only. Then we successively replaced one spike train from this cluster by a spike train from cluster
$1$, until in the end the set consisted only of spike trains from cluster $1$. This \emph{set transition} is parameterized by the set imbalance
\begin{equation} \label{eq:Set-Balance}
    b = \frac{N_1 - N_2}{N_1 + N_2}.
\end{equation}
which varies from $-1$ to $1$ along the way.
\begin{figure}
    \includegraphics[width=85mm]{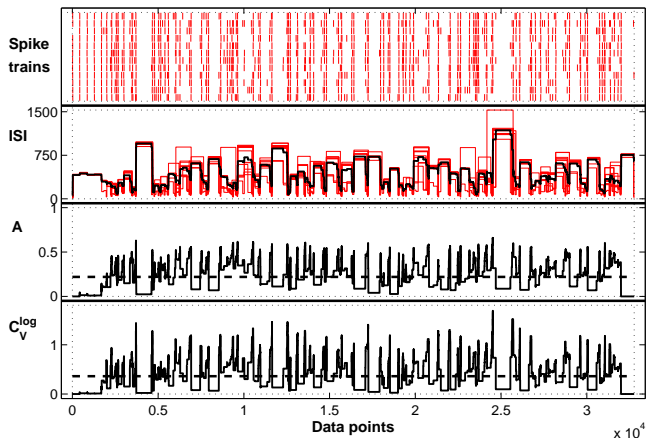}
    \caption{\abb\label{fig:HR-PA-CV-C2} (color online) Averaged bivariate ISI-distance and multivariate ISI-diversity for $N = 12$ spike trains all from cluster $2$ corresponding
        to a set imbalance of $b = -1$. In this case the ISI-measures are $D_I^a = 0.21$ and $D_I^m = 0.36$.}
\end{figure}

As a first example, Fig. \ref{fig:HR-PA-CV-C2} shows the two ISI-measures calculated for $N = 12$ spike trains that are all taken from cluster $2$ (set imbalance $b = -1$).
Similar to the examples shown in the first part and to the visualization used for the bivariate ISI-distance, parts of the spike trains that are very similar or very different
from each other are easily identified. In the beginning, small ISI-measures reflect a short transient period of almost complete synchronization. Later, the ISI-measures
highlight short-time excursions that appear when transitions between short and long ISIs are temporally offset between different spike trains.
\begin{figure}
    \includegraphics[width=85mm]{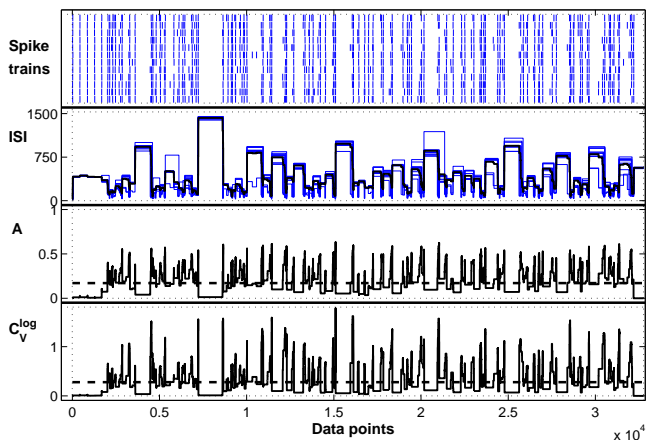}
    \caption{\abb\label{fig:HR-PA-CV-C1} (color online) Example of the ISI-measures for a set imbalance $b = 1$ with all $N = 12$ spike trains from cluster $1$.
        For this stronger connected cluster smaller ISI-measures $D_I^a = 0.17$ and $D_I^m = 0.28$ are obtained.}
\end{figure}

\begin{figure}
    \includegraphics[width=85mm]{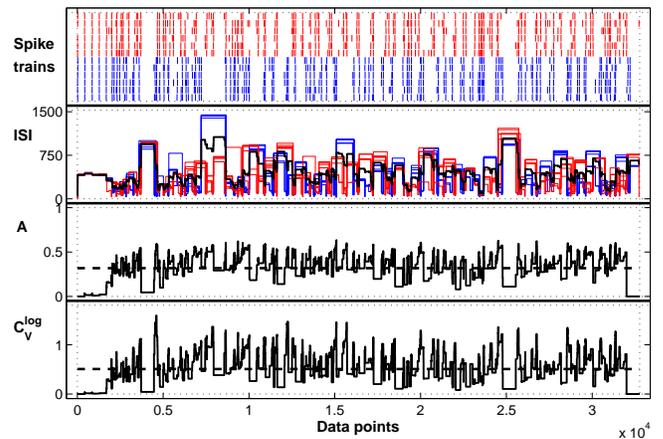}
    \caption{\abb\label{fig:HR-PA-CV-C12} (color online) ISI-measures for the balanced case ($b = 0$) with $6$ spike trains from cluster $1$ (blue)
        and $6$ spike trains from cluster $2$ (red). Correspondingly, the ISI-measures are much higher: $D_I^a = 0.32$ and $D_I^m = 0.51$.}
\end{figure}

A similar behavior can be observed for $N = 12$ spike trains from cluster $1$ (set imbalance $b = 1$, Fig. \ref{fig:HR-PA-CV-C1}). In this example, the ISI-measures are lower
than those observed for cluster $2$, which is due to the stronger coupling within cluster $1$ (cf. Fig. \ref{fig:Bi-ISI-Dist-Mat}). Finally, a set in which both clusters
contribute an equal number of $6$ spike trains (set imbalance $b = 0$, Fig. \ref{fig:HR-PA-CV-C12}) yields ISI-measures that are much higher than either of the two previous
examples.

\begin{figure}
    \includegraphics[width=85mm]{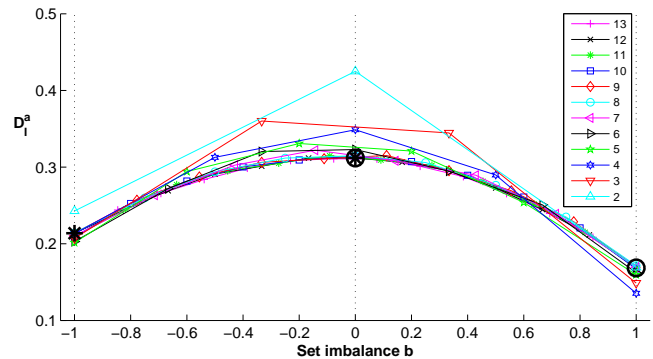}
    \caption{\abb\label{fig:Bi-ISI-Dist-vs-Set-Balance} (color online) Averaged bivariate ISI-distance versus the set imbalance (cf. Eq. \ref{eq:Set-Balance}) for randomly
        chosen set transitions and decreasing total numbers of spike trains $N$. The enlarged markers all lie on the line for $N = 12$ spike trains and correspond to the
        examples shown in Figs. \ref{fig:HR-PA-CV-C2}-\ref{fig:HR-PA-CV-C12}.}
\end{figure}

Fig. \ref{fig:Bi-ISI-Dist-vs-Set-Balance} shows the results from an application of the averaged bivariate ISI-distance to one complete group of set transitions. For each curve,
corresponding to a constant total number of spike trains $N$, a monotonic increase to the maximum value of $D_I^a$ is followed by a monotonic decrease. There is a clear
asymmetry - the values for the minimum negative set imbalance are higher than those for the maximum positive set imbalance - again reflecting the fact that the first cluster is
more strongly coupled than the second one. Different traces, corresponding to different numbers of neurons in the set, overlap considerably; only the sets with small numbers of
neurons attain mostly higher values.

The latter effect can be observed for all measures, and can easily be understood for the averaged bivariate measures. For these measures the average pairwise distance only
depends on the distance matrix; in fact, for a sufficiently high total number of spike trains $N$, it only depends on the mean distances within each of the two clusters and on
the mean distance between the two clusters (averaging over the respective submatrices yields $\langle D \rangle_1$, $\langle D \rangle_2$ and $\langle D \rangle_{12}$; compare
to Fig. \ref{fig:Bi-ISI-Dist-Mat}).

\begin{figure}
    \includegraphics[width=85mm]{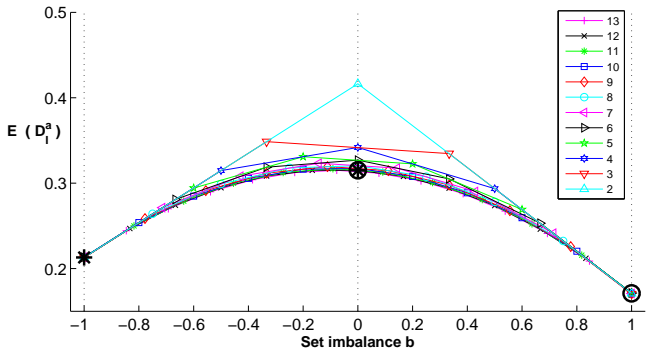}
    \caption{\abb\label{fig:Calc-Bi-ISI-Dist-vs-Set-Balance} (color online) Expected values for the averaged bivariate ISI-distance versus the set imbalance
        (cf. Eq. \ref{eq:Set-Balance}) for different total numbers of spike trains $N$. Values were calculated using Eq. \ref{eq:Multi-Bi-Distance} and the mean values of the
        submatrices shown in Fig. \ref{fig:Bi-ISI-Dist-Mat}.}
\end{figure}

For each total number of spike trains and each set imbalance, the number of pairs $N_{p1}$ and $N_{p2}$ within each cluster, the number of pairs $N_{p12}$ between the two
clusters and the total number of pairs $N_p$ can be combined with the respective mean distances to yield the expected distance
\begin{equation} \label{eq:Multi-Bi-Distance}
    E (D_X) = \\
        \frac{N_{p1} \langle D_X \rangle_1 + N_{p2} \langle D_X \rangle_2 + N_{p12} \langle D_X \rangle_{12}}{N_{p}}
\end{equation}
with $X$ representing any of the bivariate measures. For the averaged bivariate ISI-distance the expected values $E (D_I^a)$ are shown in Fig.
\ref{fig:Calc-Bi-ISI-Dist-vs-Set-Balance}. Note that for normalized set imbalances $\pm 1$ always the mean intra-cluster distances $\langle D \rangle_1 = 0.17$ and $\langle D
\rangle_2 = 0.21$ (cf. Fig. \ref{fig:Bi-ISI-Dist-Mat}) are obtained independently of the total spike train number.

For a given pairwise distance matrix with given intra- and inter-cluster distances, the value $E (D_X)$ only depends on the three coefficients. For a small total number of spike
trains there are fewer intra-cluster than inter-cluster pairs, thus the distances tend to be higher. However, for a constant set imbalance the ratio $N_{p12} / ( N_{p1} +
N_{p2})$ of inter-cluster and intra-cluster coefficients decreases as the total number of spike trains $N$ increases. With $N$ approaching infinity the ratio converges to $1$
from above, and thus the asymptotic curve for the expected value connects the marked points in Fig. \ref{fig:Calc-Bi-ISI-Dist-vs-Set-Balance} given by $\langle D_X \rangle_2$,
$\langle D_X \rangle_1/4 + \langle D_X \rangle_2/4 + \langle D_X \rangle_{12}/2$ (the relative amounts of the three submatrices marked in Fig. \ref{fig:Bi-ISI-Dist-Mat} for an
infinitively large matrix) and $\langle D_X \rangle_1$ (from left to right). However, since in reality the two intra-cluster submatrices and the inter-cluster submatrix are
non-uniform, the averaged bivariate distances for a given random set transition also depend on the standard deviation of the intra- and inter-cluster distances. Although this
kind of reasoning is only directly applicable to the bivariate measures, the observed behavior for the multivariate measures is very similar. Thus, in general the synchrony
within a small non-homogeneous neuronal population is dependent not only on the relative mixture of clusters, but also on the total number of neurons. This fact is very
important for multi-unit recordings, where typically only a relatively small number of neurons can be isolated, and where the fraction and the identity of the neurons coding for
the presented stimulus at that moment in time is not known.

In order to numerically evaluate the performance in distinguishing different sets, we randomly selected $r = 500$ set transitions for each $N = 2,...,13$, and applied all
measures to all sets. In this way for each total number of spike trains and every set imbalance we obtained a distribution of $500$ values. Then, for every set transition we
looked at the distributions of measure values for adjacent set imbalances only, and quantified their pairwise difference using Kolmogorov-Smirnov discrimination values
\citep{Stuart99}.

\begin{figure}
    \includegraphics[width=85mm]{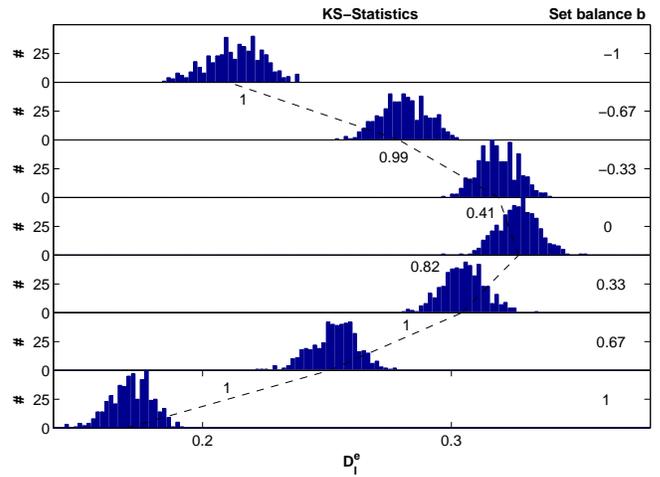}
    \caption{\abb\label{fig:Bi-ISI-Dist-Histos} Averaged bivariate ISI-distance $D_I^a$: Distributions for a constant total number of $N=6$ spike trains. The set transition takes
        place from top to bottom with the set imbalance increasing from $b=-1$ to $b=1$ (see right side). Histograms for adjacent normalized set imbalances
        are connected by dashed lines, and the accompanying numbers represent the value of the Kolmogorov-Smirnov statistic for the respective pair of distributions.}
\end{figure}

An example set transition for the averaged bivariate ISI-distance $D_I^a$ with a constant total number of $N=6$ spike trains is shown in Fig. \ref{fig:Bi-ISI-Dist-Histos}. In
this case the first and the last two pairs of distributions are disjunct, yielding Kolmogorov-Smirnov statistics equal to $K = 1$, whereas the intermediate distributions
overlapped and correspondingly the Kolmogorov-Smirnov statistics attain values smaller than one.

\begin{figure}
    \includegraphics[width=85mm]{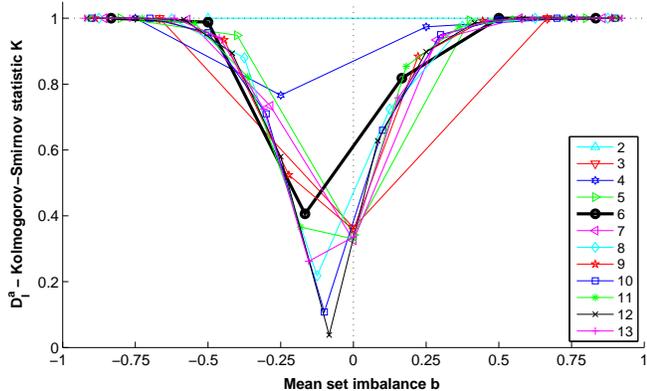}
    \caption{\abb\label{fig:Bi-ISI-Dist-KS-Stat} (color online) Kolmogorov-Smirnov statistic $K$ for the averaged bivariate ISI-distance $D_I^a$ and different total numbers
        of spike trains. The thick line corresponds to the distributions shown in Fig. \ref{fig:Bi-ISI-Dist-Histos}. Values are plotted against the mean set imbalance of the
        respective two distributions. For this measure the set separation is $S_S = 0.54$.}
\end{figure}

\begin{figure}
    \includegraphics[width=85mm]{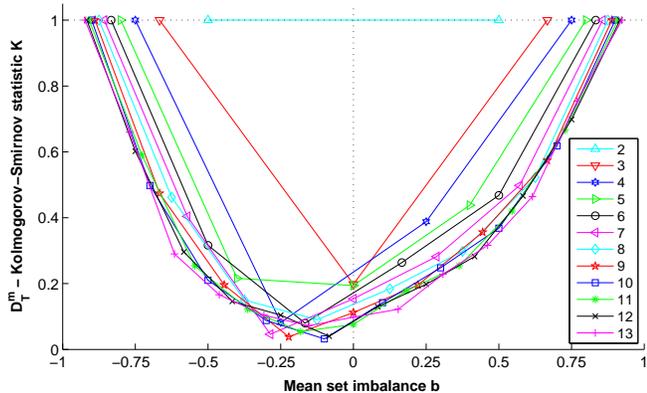}
    \caption{\abb\label{fig:Multi-Tiesinga-Dist-KS-Stat} (color online) Same as Fig. \ref{fig:Bi-ISI-Dist-KS-Stat} but this time for the Tiesinga dissimilarity $D_T^m$ which
        yields a set separation of $S_S = 0.27$.}
\end{figure}

In  Figs. \ref{fig:Bi-ISI-Dist-KS-Stat} and \ref{fig:Multi-Tiesinga-Dist-KS-Stat} an overview of the Kolmogorov-Smirnov statistic for all constant total numbers of spike trains
is shown for the averaged bivariate ISI-distance $D_I^a$ and the Tiesinga dissimilarity $D_T^m$, respectively. Most curves exhibit a perfect distinction (maximum value $K = 1$)
for very large absolute imbalances, whereas the distributions for rather balanced sets of spike trains overlap considerably (smaller values of $K$). In all cases, minimum values
are obtained for slightly negative set imbalances, again reflecting the different intra-cluster coupling strengths. However, the range of set imbalances for which the test
discriminates perfectly is much broader for the averaged bivariate ISI-distance.

\begin{figure}
    \includegraphics[width=85mm]{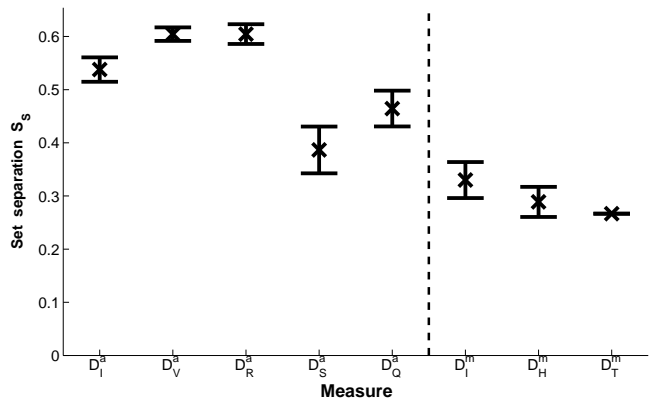}
    \caption{\abb\label{fig:KS-SetSeparation} Comparison of measures: Set separation $S_S$ defined as the fraction of tests with a Kolmogorov-Smirnov statistic of $K = 1$. The dashed line
        separates the averaged bivariate measures from the multivariate measures.   }
\end{figure}

From the ensemble of all of these traces, we defined our measure of performance, the set separation $S_S$, as the fraction of pairs of distributions that can be perfectly
distinguished, i.e., that yield a Kolmogorov-Smirnov statistic of $K = 1$ (similar results are obtained for the mean distinction, the average value over all Kolmogorov-Smirnov
statistics). In order to assess the standard error we repeated the analysis five times using $100$ realizations each time. In Fig. \ref{fig:KS-SetSeparation} we show the set
separation $S_S$ and its standard deviation for all measures. The most fundamental result is that the averaged bivariate measures are better at distinguishing different set
imbalances than the multivariate measures. The best distinctions are obtained for the averaged Victor-Purpura distance $D_V^a$ and the van Rossum distance $D_R^a$, which were
each optimized with respect to the time scale parameter, followed by the averaged bivariate ISI-distance $D_I^a$, for which no optimization was needed. The multivariate
ISI-diversity $D_I^m$ after log-transform and without any optimization performs best among the multivariate measures.

\section{\label{s:Example-Jude} Application to monkey data from a visual attention task}

\begin{figure}
    \includegraphics[width=85mm]{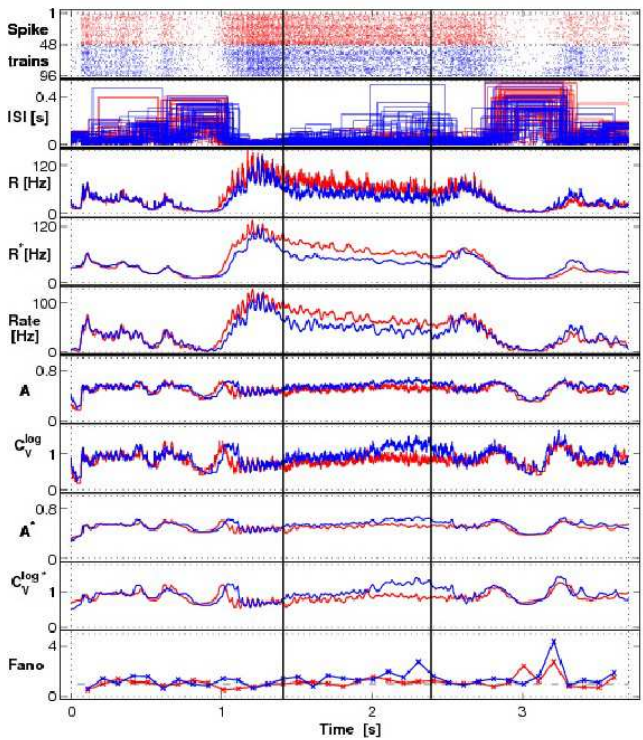}
    \caption{\abb\label{fig:JudeData} (color online) Single-unit monkey data from one neuron in area V4 during a visual attention task. In the two top panels we depict $96$ spike trains
        and their instantaneous ISI, $48$ attended (red) and $48$ unattended (blue) trials. The next three panels show firing rate estimates: the instantaneous firing rate $R$
        defined as the instantaneous average over the inverse ISIs of the individual neurons, its moving average $R^*$ as well as the smoothed mean firing rate (denoted as "Rate").
        The five bottom panels depict variability estimates based on the instantaneous bivariate average $A$ and the multivariate coefficient of variation $C_V^{log}$, their moving
        averages $A^*$ and $C_V^{log *}$, as well as the Fano factor (denoted as "Fano"). In the latter panel a horizontal dashed line marks $1$. Vertical black lines designate the
        beginning and the end of the interval in which the stimulus paused within the receptive field; the response starts approximately $400$ ms earlier when the stimulus enters
        the receptive field.}
\end{figure}

In the previous section we compared the time scale independent ISI-measures only against methods that either employ a parameter determining the time scale or are time scale
independent themselves. However, some of the standard methods used in the literature rely on dividing spike trains into disjunct or overlapping segments, which are then analyzed
using a moving window analysis. In this last section we compare the ISI-measures with such an analysis using single-unit data from one neuron in area V4 of a rhesus macaque
monkey during a visual attention task. The recordings and the task are described in Appendix \ref{App-ss:Jude-Data}, for more details see \citet{Mitchell07}. In this study
Mitchell et al. calculated the firing rate and the Fano factor (a measure of response variability: the ratio of spike count variance to mean spike count) of one neuron for
different attention conditions. The firing rate was smoothed using a Gaussian kernel with $\sigma = 12.5 ms$, whereas the Fano factor was calculated in non-overlapping $100$ ms
bins and then averaged across all visually responsive neurons for attended and unattended stimuli separately. As these two measures show (Fig. \ref{fig:JudeData}), attention
leads to an increase in firing rate and a decrease in variability (increase in reliability) during the pause period in which the attended stimulus is positioned within the
neuron's receptive field.

In Fig. \ref{fig:JudeData} we also depict the instantaneous firing rate $R (t)$ (Eq. \ref{eq:ISI-Rate}) as well as the instantaneous estimates $A (t)$ (Eq.
\ref{eq:Instantaneous-Average}) and $C_V^{log} (t)$ (Eq. \ref{eq:ISI-CV2}) for the averaged bivariate ISI-distance and the multivariate ISI-diversity, respectively. These
estimates have the highest possible resolution since they change with every occurrence of a single spike in one of the spike trains. However, in order to capture long term
trends curves can be smoothed using a moving average filter with the desired order resulting in quantities $R^* (t)$, $A^* (t)$, and $C_V^{log*} (t)$, respectively. The moving
average was applied in a manner that adapts to the local firing rate, i.e., the averaging was performed not over fixed time intervals but over fixed numbers of ISIs of the
pooled spike train. The order was set to $50$ in order to be sensitive to the same time scales that were highlighted by \citet{Mitchell07}.

A qualitative comparison between the standard methods and our proposed measures (in particular, the moving averages) reveals that the trends shown seem to be very similar.
However, the instantaneous estimates are sensitive to additional features in the spike trains, since they exhibit maximum temporal resolution. Note also that the two kinds of
approaches rely on different information. While the standard methods such as the Peri-Stimulus-Time-Histogram (PSTH) and the Fano factor capture features contained in windows of
a certain size (a parameter to be fixed beforehand), our proposed measures are time-scale-independent and parameter-free.

Furthermore, the method used above (fifth panel of Fig. \ref{fig:JudeData}), the PSTH and other estimates of firing rate based on the summed spike count are invariant to
shuffling spikes among the spike trains. They yield the same high value for a window regardless of whether the spikes are spread across the different spike trains or whether all
the spikes originate from the same spike train. In contrast, our instantaneous estimates take into account specific information from the individual spike trains by averaging
over the intervals between their previous and their next spike. This method of averaging, in which the relevant unit of the average is the spike train and not the single spike,
is closer to the actual notion of mean firing rate.

Finally, for the case of Gaussian variability of the rate across trials it can be shown that the Fano factor is affected by changes in the mean rate \citep{Chicharro09}. In
contrast, the instantaneous estimates of the ISI-distances are time-scale adaptive and thus do not depend on the mean rate.

\section{\label{s:Discussion} Discussion}

In this study we presented two extensions of the ISI-distance that measure the dissimilarity within a set of spike trains, which is typically derived either from successive
recordings from only one neuron (e.g., upon repeated presentations of the same stimulus), or from simultaneous recordings of a population of neurons. The first extension employs
the instantaneous average over the bivariate ISI-ratios whereas the second extension is multivariate and relies on the instantaneous coefficient of variation. Both of these
extensions inherit the basic properties of the ISI-distance: they are parameter free, time scale independent and easy to visualize in a time-resolved manner. Furthermore, they
are conceptually simple and computationally fast. They also yield good results even on a rather limited number of short spike trains. In the first example we dealt with only $5$
spike trains of about $60$ spikes each.

As illustrated in the first part of this study in which we analyzed \emph{in vitro} recordings of a cortical cell from a rat, the ISI-measures serve as an excellent means to
track the occurrence of firing patterns in spike trains. Deviations from synchrony can be localized in time, thus rendering these methods a good choice for time-resolved
applications on non-stationary neuronal dynamics. In the context of reliability estimations, this property allows the analyst to relate intervals of high or low synchrony to
local stimulus features.

In the second part of this study we showed that for small non-homogenous neuronal populations, the values of the averaged bivariate and the multivariate measures depend not only
on the relative mixture of clusters but also on the total number of neurons. This has important implications for the analysis of multi-unit recordings that typically access only
a rather small number of neurons, and for which the relative amount of neurons that do or do not code for the presented stimulus at that moment in time is not known.

We used a network of simulated Hindemarsh-Rose neurons to carry out a controlled comparison between the ISI-measures and six previously published methods. In this comparison the
averaged bivariate measures perform better than the multivariate measures in distinguishing different sets of spike trains. Although the averaged bivariate ISI-distance is not
the overall best performer (this is the averaged bivariate Victor-distance), it is the measure which combines sensitivity and time-resolved visualizability best. Among the
multivariate measures, the multivariate ISI-diversity is indeed the best performer. This is most likely due to the insensitivity of the other multivariate measures, the Tiesinga
and the Hunter et al. reliabilities, to the spike train of origin. Their first step is to pool the different spike trains into one cumulative spike train. Thus, for any single
spike, the information about the spike train of origin is lost, i.e., the value of the reliability methods is invariant to shuffling spikes among the spike trains. Many
different spike trains would yield the same pooled spike train (in principle even one spike train containing all of the spikes plus many empty spike trains) and thus much of the
information distinguishing different sets of spike trains taken from different clusters is lost.

This limitation does not affect the multivariate ISI-diversity, which is more sensitive to local structure within the single spike trains, since the interspike intervals are
defined by neighboring spikes of the same spike train. However, the multivariate ISI-diversity still performs not as well as the averaged bivariate measures, because the
calculation of the instantaneous mean and standard deviation across spike trains implicitly assumes that a unimodal distribution exists. Spike trains belonging to different
clusters result in a multimodal distribution which cannot be well-described by a common mean and a common standard deviation. Thus, although the multivariate ISI-diversity is a
step in the right direction, further improvements are needed for it and for all multivariate methods in order for them to compete with the averaged bivariate methods.

One difference between the ISI-measures and most of the other methods is that for the former, we neither have to divide the time series into windows of a certain fixed size, nor
need to choose a parameter that sets the time scale of analysis. In our quantitative comparison this difference was less apparent since we only showed results obtained for time
scale parameters that have already been optimized. However, in applications to real data for which there is no validated knowledge about the data and its relevant time scales,
it can be preferable to get a more objective estimate of neuronal variability by using a method that is time scale independent. Of course the computational cost and the time and
effort that is needed to find the right parameter, if there indeed is one, might also play a role. However, it is also important to keep in mind that for studies dealing with
specific questions of neuronal coding (e.g., what is the time scale over which different stimuli can be best distinguished?) the use of a parameter-based methods is preferable.
But even in such cases, it might be useful to get a time scale independent estimate of the fundamental distinguishability of the spike trains in question. This holds
particularly true for spike trains that contain different time scales such as regular spiking and bursting, since in these cases the optimum time scale can become ambiguous.

In the third part of this study we reanalyzed single-unit monkey data to compare our methods with the Fano factor, a standard method that employs a moving window technique. In
this application we qualitatively demonstrated that our instantaneous measures allow us to analyze the data at the maximum possible resolution, but that this resolution can then
easily be reduced by means of a moving average filter with the desired order. On the other hand, for measures like the Peri-Stimulus-Time-Histogram (PSTH) and the Fano factor,
which rely on windowing, maximum resolution is always limited due to the demands on the spike counting statistics.

A few of the caveats that one should take into account for the analysis of two spike trains also hold for the multiple spike train case. First, it is obvious that no measure
that results in a single number quantifying the synchrony between two or more spike trains can be adequate to deal with all kinds of potential coding schemes (e.g, time coding,
rate coding and pattern coding; cf. also \citet{Victor97}. Then, the ISI-measures, like the other measures, are not sensitive to phase lags. Thus any such phase lags should be
removed by suitably shifting the time series before computing the measures. Methods for the detection of lag synchronization in neuronal data \citep{Waddell07} as well as for
their elimination \citep{Nawrot03} have been described. Finally, for the ISI-distances as for most of the other measures relative changes in value across different sets,
stimuli, conditions, etc. are often more important than the absolute values.

Note that the multiple spike train extensions proposed in this study aim at the quantification of synchrony within one set of spike trains. For the Victor-Purpura and the van
Rossum distances, there also exists a second group of extensions designed to estimate the synchrony between two populations of neurons (\citeauthor{Aronov03},
\citeyear{Aronov03} and \citeauthor{Houghton08}, \citeyear{Houghton08}, respectively). A corresponding extension of the ISI-distance will be presented in a forthcoming study
\citep{Kreuz09b}.

In summary, due to their time-resolved definition, time scale independence and computational speed, we expect the ISI-measures to be a powerful tool for all kinds of
applications on multiple spike train synchrony and reliability. We close by pointing out once more that the Matlab source codes for calculating and visualizing the instantaneous
firing rate and the ISI-measures can be downloaded at http://inls.ucsd.edu/$\sim$kreuz/Source-Code/Spike-Sync.html).

\begin{appendix} \label{s:Appendix}

\section{\label{App-s:Theory} Analytical results for Poisson processes}

In this Section we derive analytical results for the mean and the standard deviation for the averaged bivariate ISI-distance $D_I^a$ and the ISI-diversity $D_I^m$ in the case of
multiple Poisson processes. For this process the spike count in a window follows a Poisson distribution, whereas the ISI distribution is exponential.

An exponential ISI-distribution as obtained for a Poisson process with rate $\lambda$ yields a coefficient of variation $C_V = 1$ (both standard deviation and mean are equal to
$1/ \lambda$). However, relevant for the instantaneous measures is the probability at each time instant to observe the interspike interval $x$. This probability is the product
of the length and the frequency of the interval (with both of these quantities expressed in units of the mean $1/ \lambda$), i.e.,
\begin{equation}
    P(x) = \lambda^2xe^{-\lambda x}.
\end{equation}
For the bivariate ISI-distance $D_I$ we calculate the expectation value of the instantaneous ISI-ratio $I(t)$ (Eq. \ref{eq:ISI-Ratio}) using this probability distribution for
both $x_1$ and $x_2$ \citep{Chicharro09}:
\begin{eqnarray} \label{eq:Poisson-PA-Expectation}
    E( I(t) ) = \lambda^4\int_{0}^{\infty}\int_{0}^{x_2}dx_1dx_2(1-\frac{x_1}{x_2})x_1e^{-\lambda x_1}x_2e^{-\lambda x_2}+ \\ \nonumber
        \lambda^4\int_{0}^{\infty}\int_{0}^{x_1}dx_1dx_2(1-\frac{x_2}{x_1})x_1e^{-\lambda x_1}x_2e^{-\lambda x_2} = \frac{1}{2}.
\end{eqnarray}
Note that the result is independent of the rate. According to Eqs. \ref{eq:Instantaneous-Average} and \ref{eq:Multi-ISI-distance} the same expectation value is obtained for the
averaged bivariate ISI-distance $D_I^a$ and the instantaneous average $A(t)$.

Correspondingly, for the ISI-diversity $D_I^m$ we calculate the expectation value of the instantaneous coefficient of variation $C_V(t)$ which we get from the mean and the
standard deviation of the above distribution (Eq. \ref{eq:ISI-CV}).
\begin{figure}
    \includegraphics[width=85mm]{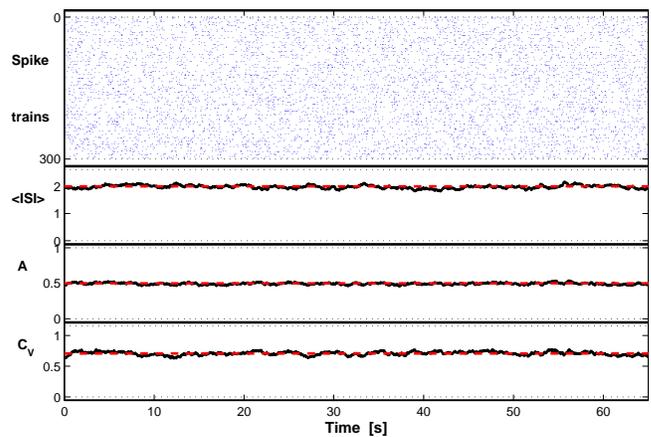}
    \caption{\abb\label{fig:PoissTest} (color online) Multiple Poisson spike trains ($N = 300$) each with the same rate $\lambda = 1$. Below are shown the instantaneous mean ISI,
        the instantaneous average  bivariate ISI-ratio $A$ and the multivariate coefficient of variation $C_V$. The dashed red lines mark the expectation values for the
        instantaneous variability estimates calculated according to Eqs. \ref{eq:Poisson-PA-Expectation} and \ref{eq:Poisson-CV-Expectation}.}
\end{figure}

The mean of this distribution is
\begin{equation} \label{eq:Poisson-Mean-ISI-Expectation}
    \mu = \int_{0}^{\infty} dx \lambda^2x^2e^{-\lambda x}= \frac{2}{\lambda}
\end{equation}
whereas its standard deviation is
\begin{equation}
    \sigma = \sqrt{<x^2>-\mu^2} = \frac{\sqrt{2}}{\lambda}.
\end{equation}

\noindent This yields the expectation value for the coefficient of variation
\begin{equation} \label{eq:Poisson-CV-Expectation}
    E ( C_V(t) ) = \frac{1}{\sqrt{2}}.
\end{equation}
which is also independent of the rate but in fact considerably smaller than $1$.

In Fig. \ref{fig:PoissTest} we show the instantaneous analysis for $300$ Poisson spike trains each with the same rate $\lambda = 1$. The expected values $<ISI> = 2$ (Eq.
\ref{eq:Poisson-Mean-ISI-Expectation} with $\lambda = 1$), $A = 0.5$ and $C_V = 1/\sqrt{2}$ are marked by dashed red lines.

%
%
%
\section{\label{App-s:Data} Data}

All recordings and simulations were performed prior to and independently from the design of this study.

\subsection{\label{App-ss:Data-Stellate-Cells} In vitro recordings of a cortical cell}

In Section \ref{s:Illustration-Real-Data} the ISI-measures are illustrated using \emph{in vitro} whole-cell recordings taken from a cortical cell from the layer 2 of the medial
entorhinal cortex of a young Long-Evans rat. In these experiments (conducted as approved by the University of California at San Diego Institutional Animal Care and Use
Committee) cortical cells were selected from a $400$ micron slice preparation by their superficial position, as well as particular characteristics of their electrophysiological
responses to long current steps \citep[cf.][]{Haas02}. Intracellular signals were amplified, low pass filtered, and digitized at $10$ kHz via software created in LabView
(National Instruments, Austin, TX, USA). Inputs were delivered as synaptic conductances through a linux-based dynamic clamp \citep{Dorval01}. Inputs were comprised of synaptic
inputs added to an underlying DC depolarization. The amplitude of the DC depolarization was adjusted for each cell to elicit a spike rate of $5-10$ Hz. Synaptic inputs were of
the form $I_{syn} = G_{syn}S(V_m-V_{syn})$ where $S$ followed the differential expression $dS/dt = \alpha (1-S) - \beta S$; $\alpha=500/ms$, $\beta = 250/ms$, $V_{syn} = 0$
$mV$. $G_{syn}$ was tailored for each cell to be peri-threshold and elicit a spike with probability close to $50\%$. Synaptic events were delivered for $10$ seconds at irregular
intervals that were taken from recordings of spike times in response to steady DC depolarization alone. Inputs were temporally scaled resulting in firing rates that were roughly
matched to the elicited firing rate of the neuron, about one third slower and about twice as fast, with the latter covering only about half the length of the recording time. For
each scaling five repetitions were analyzed.

\subsection{\label{App-ss:Data-Hindemarsh-Rose} Hindemarsh-Rose simulations}

The spike trains of the controlled configuration used in Section \ref{s:Comparison-HR} were generated using time series extracted from a network of Hindemarsh-Rose neurons
\citep{Hindmarsh84}. This network was originally designed to analyze semantic memory representations using feature-based models. Details of the network architecture and the
implementation of the feature coding can be found in \citet{Morelli05}. The clustering properties of these data were detailed in \citet{Kreuz07c}.

In short, the state of the neuron $n$ was determined by three dimensionless first-order differential equations describing the evolution of the membrane potential $X_n$, the
recovery variable $Y_n$, and a slow adaptation current $Z_n$,
\begin{eqnarray} \label{eq:Hindemarsh-Rose}
    \dot{X_n} &=& Y_{n} - X_n^3 + 3X_n^{2} - Z_n + I_n + \alpha_n - \beta_n \\
    \dot{Y_n} &=& 1 - 5X_n^2 - Y_n \\
    \dot{Z_n} &=& 0.006 [4(X_n-1.6)-Z_n],
\end{eqnarray}
where
\begin{equation} \label{eq:HR-impulse-current}
    \alpha_n = \sum_{m=1}^{F(U-1)} w_{nm} A_m
\end{equation}
and
\begin{equation} \label{eq:HR-local-inhibition}
    \beta_n = \frac{1}{F-1} \sum_{k=1}^{F-1} A_k
\end{equation}
are the weighted inter-modular and intra-modular activities, respectively, which were dependent on the synaptic connection weights $w_{nm}$ and the neuronal activity variables
$A_n$. The external input $I_n$ was chosen such that the Hindemarsh-Rose neurons were operating in a chaotic regime.

The network consisted of $128$ Hindemarsh-Rose neurons belonging to $U = 16$ different modules with $F = 8$ neurons each. In a learning stage, input memory patterns were stored
by updating the synaptic connection weights $w_{nm}$ between different neurons using a Hebbian mechanism based on the activity variables $A_n$. A neuron was considered active
whenever its membrane potential $X$ exceeded the threshold value $\widehat{X} = 0$ and its activity is coded by the variable $A_n = \theta(X_n - \widehat{X})$, where $\theta$ is
the Heavyside function with $\theta(x) = 1$ if $x \geq 0$ and $\theta(x) = 0$ if $x < 0$. For this study we restricted ourselves to $26$ time series $X_n$ extracted during the
retrieval stage in which the learned connection weights were kept constant. According to their coding properties regarding the retrieval of only two distinguished memory
patterns, they belonged to two principal clusters: $13$ of the neurons coded for pattern $1$ only, and $13$ coded for pattern $2$ only. The respective time series were labelled
by ``1'' and ``2'' followed by an index letter. The numerical integration was done using a fourth-order Runge-Kutta integration with a fixed step-size of $0.05$ (arbitrary time
units). The length of the time series was $32768$ data points. The threshold for spike detection was chosen as the arithmetic average over the minimum and maximum value of the
respective time series.

\subsection{\label{App-ss:Jude-Data} Monkey data from a visual attention task}

In Section \ref{s:Example-Jude} our instantaneous measures are compared against standard methods of rate and variability estimation. This comparison is carried out on
single-unit data from a neuron in area V4 of a rhesus macaque monkey that were recorded during an attention-demanding multiple-object tracking task. These data were described in
more detail in \citet{Mitchell07}. In short, at the beginning of the recording session, the receptive field of the neuron was mapped using subspace reverse correlation on $60$
Hz Gabor stimuli. The monkey began each trial by fixating a central point for $200$ ms and then maintained fixation through the trial. Four identical Gabor stimuli appeared
outside the neuron's receptive field at equally eccentric positions separated by $90 ^{\circ}$. Two stimuli were briefly identified as targets before all stimuli moved along
independent trajectories at approximately $10^{\circ}$/s for $950$ ms, placing them at a new set of equally spaced locations, one of which was within the receptive field. The
stimuli paused for $1000$ ms before moving to another set of locations and stopping. In the end of the trial the fixation point disappeared, and the monkey made a saccade to
each target. Trials were categorized as either attended or unattended depending on the type of stimulus that entered and paused in the receptive field. For each of the two
attention conditions eight novel sets of stimulus trajectories were generated and each repeated six times resulting in a total number of $96$ trials. The data can be downloaded
from J. F. Mitchell's webpage: http://www.snl.salk.edu/$\sim$jude/sfn2008/index.html.
%

\section{\label{App-s:Existing-Methods} Previously published measures of spike train distance}

In Section \ref{s:Comparison-HR} we use a controlled configuration to compare the performance of the average bivariate ISI-distance $D_I^a$ and the multivariate ISI-diversity
$D_I^m$ to four averaged bivariate and two multivariate measures. The four averaged bivariate measures $D_V^a$, $D_R^a$, $D_S^a$ and $D_Q^a$ are based on the spike train
distances introduced by \citet{Victor96} and \citet{VanRossum01}, the similarity measure proposed by \citet{S_Schreiber03}, and event synchronization \citep{QuianQuiroga02b},
respectively. Multivariate measures comprise the reliabilities $D_H^m$ by \citet{Hunter98} and $D_T^m$ by \citet{Tiesinga04}.

The four measures $D_V^a$, $D_R^a$, $D_S^a$ and $D_H^m$ rely on a parameter that sets the time scale of the analysis. Following \citet{Kreuz07c} these parameters were varied
over several logarithmic decades, with four equidistant values within each decade, and then optimized on the Hindemarsh-Rose data, this time with respect to the present task
(see Section \ref{s:Comparison-HR}). For the averaged bivariate methods the time scales yielding the best performance were identical to those found for the bivariate measures in
\citet{Kreuz07c}, where the performance in reproducing the clustering of the same Hindemarsh-Rose network was evaluated. This is not surprising since both of these tasks rely on
the distinction of the same spike trains.

\subsection{\label{App-ss:Bivariate-measures} Averaged bivariate measures}

For this class of measures, to which the averaged bivariate ISI-distance belongs as well, synchrony is defined as the average value over all pairs of spike trains. The
underlying bivariate measures are described below. To simplify notation the two spike trains are here labelled $x$ and $y$.

\subsubsection{\label{App-sss:Victor} Victor-Purpura spike train distance}

The spike train distance $D_V$ introduced in \citet{Victor96} defines the distance between two spike trains in terms of the minimum cost of transforming one spike train into the
other by using just three basic operations: spike insertion, spike deletion and spike movement. While the cost of insertion or deletion of a spike is set to one, the cost of
moving a spike by some interval is the only parameter of the method, and sets the time scale of the analysis. For zero cost, the distance is equal to the difference in spike
counts, for high costs, the distance approaches the number of non-coincident spikes, as it becomes more favorable to delete all non-coincident spikes than to shift them. Thus,
by increasing the cost, the distance is transformed from a rate distance to a timing distance. Optimization of the cost on the Hindemarsh-Rose data yielded an intermediate value
of $0.01$/time unit.

\subsubsection{\label{App-sss:vanRossum} Van Rossum spike train distance}

A second spike train distance was introduced in \citet{VanRossum01}. In this method, each spike is convolved with an exponential function with time constant $\tau_R$. From the
convolved waveforms $\tilde{x}(t)$ and $\tilde{y}(t)$, the van Rossum distance $D_R$ can be calculated as
%
%
\begin{equation} \label{eq:RossumDistance}
    D_R (\tau_R) = \frac{1}{\tau_R} \int_0^\infty [ \tilde{x}(t) - \tilde{y}(t) ]^2 dt
\end{equation}
Since the post-synaptic currents triggered by single spikes are well fit by exponentials, the van Rossum distance estimates the difference in the effect of the two trains on the
respective synapses. In this method, the time constant $\tau_R$ of the exponential as the parameter that sets the time scale. Following optimization it is set to $\tau_R =
177.83$ time units.

\subsubsection{\label{App-sss:Schreiber} Schreiber et al. similarity measure}

In this correlation-based approach (\citeauthor{S_Schreiber03}, \citeyear{S_Schreiber03}; cf. also \citeauthor{Haas02}, \citeyear{Haas02}) each spike train is convolved with a
Gaussian filter of width $\sigma_S$ to form $\tilde{x}'$ and $\tilde{y}'$ before cross correlation and normalization:
\begin{equation} \label{eq:SchreiberCorrelation}
    C_S (\sigma_S) = \frac{\tilde{x}' \cdot \tilde{y}'}{|\tilde{x}'||\tilde{y}'|}.
\end{equation}
To derive a normalized measure of spike train dissimilarity the measure is inverted according to $D_S = 1-C_S$. The width of the convolving filter $\sigma_S$ sets the time scale
of interaction between the two spike trains. Optimization resulted in a value of $\sigma_S = 31.62$ time units.

\subsubsection{\label{App-sss:Event-Synchronization} Event synchronization}

Event synchronization (\citeauthor{QuianQuiroga02b}, \citeyear{QuianQuiroga02b}; cf. also \citeauthor{Kreuz07a}, \citeyear{Kreuz07a}) works as a coincidence detector quantifying
the level of synchrony from the number of quasi-simultaneous appearances of spikes. In contrast to the measures introduced above, this method is parameter- and scale-free, since
the time lag $\tau_{ij}$ up to which two spikes $t_i^x$ and $t_j^y$ are considered to be synchronous is adapted to the local spike rates:
\begin{equation} \label{eq:Event-MaxDist}
    \tau_{ij} = \min \{t_{i+1}^x - t_i^x, t_i^x - t_{i-1}^x,
                           t_{j+1}^y - t_j^y, t_j^y - t_{j-1}^y\}/2.
\end{equation}
The function $c(x|y)$ counts the appearances of a spike in $x$ shortly after a spike in $y$:
\begin{equation} \label{eq:Event-Count}
    c (x|y) = \sum_{i=1}^{M_x} \sum_{j=1}^{M_y} J_{ij},
\end{equation}
where
\begin{equation} \label{eq:Event-Synchronicity}
    J_{ij} = \begin{cases}
                      1     & {\rm if} ~~ 0 < t_i^x - t_j^y \leq \tau_{ij} \cr
                      1/2   & {\rm if} ~~ t^x_i = t^y_j \cr
                      0     & {\rm otherwise}.
                  \end{cases}
\end{equation}
With $c(y|x)$ defined accordingly, event synchronization can be written as
\begin{equation} \label{eq:Event-Synchro}
    Q = \frac {c (y|x) + c (x|y)} {\sqrt{M_x M_y}}.
\end{equation}
A suitably normalized distance is obtained by inversion $D_Q=Q-1$, with $D_Q = 0$ if and only if all spikes coincide.

\subsection{\label{App-ss:Multivariate-measures} Multivariate measures}

The first step for the \citet{Hunter98} and the \citet{Tiesinga04} reliability measures is to pool the spike times of all neurons together into one set
${t^P_1,...,t^P_i,...,t^P_{M^P}}$ with $M^P$ denoting the total number of spikes. The interspike intervals of this pooled spike train are labelled as $p_{\mathrm {ISI}}^i=
t^P_{i+1}-t^P_i$.

\subsubsection{\label{App-sss:Hunter} Hunter et al. reliability}

Like the van Rossum distance the multivariate reliability measure $D_H^m$ introduced by \citet{Hunter98} takes distances between spikes into account by convolving each spike of
the pooled spike train with an exponential function with time constant $\tau_H$. Then it evaluates the variance $V^P$ of the resulting continuous function, which is not bounded
and expected to be high for synchronous and low for independent spike trains. The normalization proposed in \citet{Hunter98} relies on some general assumptions that are not
fulfilled in our case, e.g., the same number of spikes $M$ in all spike trains. Thus, we normalized the variance by the value $V^P_{max}$ obtained for the set of identical spike
trains that is generated by reproducing $N$ times the single spike train of the set that gives the maximum variance value, i.e., $V^P_{max} = N^2 \max_{n=1,...,N} \{V^n\}$ where
$V^n$ denotes the variance of the train from neuron $n$. We then take the complement to $1$ in order to get a measure of dissimilarity:
\begin{equation} \label{eq:HunterDistance}
    D_H^m = 1 - \frac{V^P}{V^P_{max}}.
\end{equation}
The time constant $\tau_H$ of the exponential sets the time scale. Best results were obtained for $\tau_H = 56.23$ time units.

\subsubsection{\label{App-sss:Tiesinga} Tiesinga reliability}

The first step for the measure introduced in \citet{Tiesinga04} is to calculate the coefficient of variation for the pooled spike train
\begin{equation} \label{eq:PooledCV}
    C_V^P = \frac{\sigma (p_{\mathrm {ISI}}^i)}{\langle {p_{\mathrm {ISI}}^i} \rangle_i}.
\end{equation}
For an asynchronous ensemble each neuron fires uncorrelated to the other. The spike time histogram is therefore flat. For a homogeneous Poisson spike train, the coefficient of
variation $C_V$ of a single neuron's interspike interval distribution will be asymptotically equal to one. The superposition of independent Poisson processes is also a Poisson
process with a $C_V^P$ equal to one as well. The variance in the $C_V^P$ value across different realizations of the ensemble depends only on the number of spikes $M^P$ in the
pooled spike train, not on the number of neurons $N$ in the ensemble. By normalizing the $C_V^P$ to the analytically computed value for a perfectly synchronous ensemble
\citep[for details see][]{Tiesinga04} and inverting one obtains the measure of dissimilarity
\begin{equation} \label{eq:TiesingaDistance}
    D_T^m = 1 - \frac{C_V^P - 1}{\sqrt{N}}.
\end{equation}
that is sensitive to the precision of the ensemble discharge as well as to the degree of coincidence. This method is free of parameters.

\end{appendix}

\vspace{1cm}

 \begin{thanks}
    \textbf{Acknowledgements:}
     We thank Alice Morelli for the Hindemarsh-Rose data, and Jude Mitchell and John Reynolds for the monkey data and the Matlab source codes for the firing rate and the
     Fano factor. We further thank Tim Gentner (and all members of his lab), Peter Grassberger, Antonio Politi and Alessandro Torcini for useful discussions. TK has been supported
     by the Marie Curie Individual Outgoing Fellowship ``STDP'', project No 040576. RGA acknowledges grant BFU2007-61710 of the Spanish Ministry of Education and Science.
     DC was supported by the grant 2008FI-B 00460 of the ``Generalitat de Catalunya'' and the European Social Fund.
 \end{thanks}


\begin{thebibliography}{30}

\bibitem[{Aronov (2003)}]{Aronov03}
Aronov D, Reich DS, Mechler F, Victor JD. Neural coding of spatial phase in V1 of the macaque monkey. J Neurophysiol, 2003;89:3304-27.

\bibitem[{Brown et~al.(2004)}]{Brown04}
Brown EN, Kass RE, Mitra PP. Multiple neural spike train data analysis: state-of-the-art and future challenges. Nature Neurosci, 2004;7:456-61.

\bibitem[{Buzsaki and Draghun(2004)}]{Buzsaki04}
Buzsaki G, Draghun A. Neuronal Oscillations in Cortical Networks. Science, 2004;304:1926-29.

\bibitem[{Chicharro et~al.(2009)}]{Chicharro09}
Chicharro D, Kreuz T, Andrzejak RG. On the specificity of spike train reliability measures. In preparation.

\bibitem[{Dorval et~al.(2001)Dorval, Christini, and White}]{Dorval01}
Dorval AD, Christini DJ, White JA. Real-time linux dynamic
  clamp: A fast and flexible way to construct virtual ion channels in living
  cells. Annals of Biomedical Engineering, 2001;29:897-907.

\bibitem[{Gerstein et~al.(2001)}]{Gerstein01}
Gerstein GL, Kirkland KL. Neural assemblies: technical issues, analysis and modeling. Neural Networks, 2001;14:589-98.

\bibitem[{Haas and White(2002)}]{Haas02}
Haas JS, White JA. Frequency selectivity of layer II stellate
  cells in the medial entorhinal cortex. J Neurophysiol, 2002;88:2422-29.

\bibitem[{Haas et~al.(2009)}]{Haas09}
Haas JS, Kreuz T, Torcini A, Politi A, Abarbanel HDI. Resonance and rate maintenance in spiking neurons driven with strong inputs. Submitted, 2009.

\bibitem[{Hindmarsh and Rose(1984)}]{Hindmarsh84}
Hindmarsh JL, Rose RM. A model of neuronal bursting using three
  coupled first order differential equations. Proc R Soc London B, 1984;221:87-102.

\bibitem[{Hopkins et~al.(2000)}]{Hopkins00}
Hopkins WG. Measures of Reliability in Sports Medicine and Science. Sports Medicine, 2000;30:1-15.

\bibitem[{Houghton and Sen(2008)}]{Houghton08}
Houghton C, Sen K. A new multineuron spike train metric. Neural Computation, 2008;20:1495-1511.

\bibitem[{Hunter et~al.(1998)}]{Hunter98}
Hunter JD, Milton G, Thomas PJ, Cowan JD. Resonance effect
  for neural spike time reliability. J Neurophysiol, 1998;80:1427-38.

\bibitem[{Kreiman et~al.(2000)}]{Kreiman00}
Kreiman G, Krahe R, Metzner W, Koch C, Gabbiani F.
  Robustness and variability of neuronal coding by amplitude-sensitive afferents in the weakly electric fish Eigenmannia. J Neurophysiol, 2000;84:189-204.

\bibitem[{Kreuz et~al.(2007)}]{Kreuz07c}
Kreuz T, Haas JS, Morelli A, Abarbanel HDI, Politi A.
  Measuring spike train synchrony. J Neurosci Methods, 2007;165:151-61.

\bibitem[{Kreuz et~al.(2007b)}]{Kreuz07a}
Kreuz T, Kraskov A, Andrzejak RG, Mormann F, Lehnertz K, Grassberger P.
  Measuring synchronization in coupled model systems: A
  comparison of different approaches. Phys D, 2007;225:29-42.

\bibitem[{Kreuz et~al.(2009)}]{Kreuz09b}
Kreuz T, Chicharro D, Andrzejak RG.
  Measuring population spike train synchrony. In preparation.

\bibitem[{Mainen and Sejnowski(1995)}]{Mainen95}
Mainen Z, Sejnowski TJ. Reliability of spike timing in neocortical
  neurons. Science, 1995;268:1503-6.

\bibitem[{Mitchell et~al.(2007)}]{Mitchell07}
Mitchell JF, Sundberg KA, Reynolds JH. Differential attention-dependent response modulation across cell classes in macaque visual area \textsc{V}4. Neuron, 2007;55:131-41.

\bibitem[{Morelli et~al.(2005)}]{Morelli05}
Morelli A, Grotto RL, Arecchi FT. A feature-based model of semantic memory: \textsc{T}he importance of being chaotic. Lecture Notes in Computer Science, 2005;3704:328-37.

\bibitem[{Nawrot(2003)}]{Nawrot03}
Nawrot MP, Aertsen A, Rotter S. Elimination of response latency variability in neuronal spike trains. Biol Cybern, 2003;88:321-34.

\bibitem[{{Quian Quiroga} et~al.(2002){Quian Quiroga}, Kreuz, and
  Grassberger}]{QuianQuiroga02b}
{Quian Quiroga} R, Kreuz T, Grassberger P, 2002. Event synchronization:
  \textsc{A} simple and fast method to measure synchronicity and time delay
  patterns. Phys Rev E, 2002;66:041904.

\bibitem[{Schreiber et~al.(2003)}]{S_Schreiber03}
Schreiber S, Fellous JM, Whitmer JH, Tiesinga PHE, Sejnowski TJ.
  A new correlation-based measure of spike timing reliability.
  Neurocomputing, 2003;52:925-31.

\bibitem[{Schwartz(2006)}]{Schwartz06}
Schwartz O, Pillow JW, Rust NC, Simoncelli EP. Spike-triggered neural characterization. Journal of Vision, 2006;88:484-507.

\bibitem[{Stuart et~al.(1999)}]{Stuart99}
Stuart A, Ord JK, Arnold S. Kendall's advanced theory of statistics. Vol. 2A: Classical inference and the linear model. Hodder Arnold: London, 1999.

\bibitem[{Tiesinga(2004)}]{Tiesinga04}
Tiesinga PHE. Chaos-induced modulation of reliability boosts output
  firing rate in downstream cortical areas. Phys Rev E, 2004;69:031912.

\bibitem[{Tiesinga et~al.(2008)}]{Tiesinga08}
Tiesinga PHE, Fellous JM and Sejnowski TJ. Regulation of spike timing in visual cortical circuits. Nature Rev Neurosci, 2008;9:97-107.

\bibitem[{{van Rossum}(2001)}]{VanRossum01}
{van Rossum} MCW. A novel spike distance. Neural Comput, 2001;13:751-63.

\bibitem[{Victor and Purpura(1996)}]{Victor96}
Victor JD, Purpura KP. Nature and precision of temporal coding in
  visual cortex: A metric-space analysis. J Neurophysiol, 1996;76:1310-26.

\bibitem[{Victor and Purpura(1997)}]{Victor97}
Victor JD, Purpura KP. Metric-space analysis of spike trains: theory, algorithms and application. Network: Comput Neural Syst, 1997;8:127-64.

\bibitem[{Waddell et~al.(2007)}]{Waddell07}
Waddell J, Dzakpasu R, Booth V, Riley B, Reasor J, Poe G, Zochowski M. Causal entropies—A measure for determining changes in the temporal organization of neural systems. J
Neurosci Methods, 2007;162:320-32.


\end{thebibliography}

\bibliographystyle{elsart-harv}

\end{document}